\documentclass[showpacs,pra,a4paper,floatfix,twocolumn]{revtex4}
\usepackage{graphicx}
\usepackage{bm}
\usepackage{dsfont}
\usepackage{amsmath}
\usepackage{amssymb}
\newcommand{\ket}[1]{\ensuremath{| #1\rangle}}
\newcommand{\bra}[1]{\ensuremath{\langle #1 |}}

\begin{document}

\title{Light propagation in chiral media}

\author{Robert \surname{Fleischhaker}}
\email{robert.fleischhaker@mpi-hd.mpg.de}

\author{J\"org \surname{Evers}}
\email{joerg.evers@mpi-hd.mpg.de}

\affiliation{Max-Planck-Institut f\"ur Kernphysik, Saupfercheckweg 1, 
D-69117 Heidelberg, Germany}

\date{\today}

\allowdisplaybreaks[2]

%%%%%%%%%%%%%%%%%%%%%%%%%%%%%%%%%%%%%%%%%%%%%%%%%%%%%%%%%%%%%%%%%
%%%%%%%%%%%%%%%%%%%%%%%%%%%%%%%%%%%%%%%%%%%%%%%%%%%%%%%%%%%%%%%%%
%%%%%%%%%%%%%%%%%%%%%%%%%%%%%%%%%%%%%%%%%%%%%%%%%%%%%%%%%%%%%%%%%
\begin{abstract}
Light propagation in chiral media is discussed. We derive the wave equations for a probe pulse propagating through a chiral medium, and solve them analytically in Fourier space using the slowly varying envelope approximation. Our analysis reveals the influence of the different medium response coefficients on the propagation dynamics. Applying these results to a specific example system, we show that chiral interactions already become important at experimentally accessible parameter ranges in dilute vapors. The chirality renders the propagation dynamics sensitive to the phase of the applied fields, and we show that this phase-dependence enables one to control the pulse evolution during its propagation through the medium. Our results demonstrate that the magnetic field component of a probe beam can crucially influence the system dynamics even if it couples to the medium only weakly. 
\end{abstract}


%42.50.Gy
%Effects of atomic coherence on propagation, absorption, 
%and amplification of light; electromagnetically induced 
%transparency and absorption 
%
%42.65.Sf
%Dynamics of nonlinear optical systems; optical 
%instabilities, optical chaos and complexity, and optical 
%spatio-temporal dynamics
%
%42.65.An
%Optical susceptibility, hyperpolarizability 

\maketitle

%%%%%%%%%%%%%%%%%%%%%%%%%%%%%%%%%%%%%%%%%%%%%%%%%%%%%%%%%%%%%%%%%
%%%%%%%%%%%%%%%%%%%%%%%%%%%%%%%%%%%%%%%%%%%%%%%%%%%%%%%%%%%%%%%%%
%%%%%%%%%%%%%%%%%%%%%%%%%%%%%%%%%%%%%%%%%%%%%%%%%%%%%%%%%%%%%%%%%
\section{\label{secintro}Introduction}
Quantum optics describes the interaction of matter and light at the quantum level~\cite{scullybook,loudonbook}. In many cases, the underlying principles can already be understood from a fairly simple analysis, focusing on single atoms interacting with the electric field component of the driving laser fields. But recently, motivated by the aim of achieving a negative index of refraction in atomic media~\cite{negindex1,negindex2,negindex3,negindex4,negindex5}, a qualitatively different class of setups received a considerable amount of interest. The key feature of these setups is that a probe beam interacts with the medium both with its electric and its magnetic field component at the same time. Only if both the magnetic and the electric response are suitable, a negative refractive index (NRI) can be achieved. Related to the need for a high magnetic response, these systems have two more distinct features in common. First, they operate at a high density, such that the different atoms in the medium are no longer independent~\cite{ddQFT1,ddQFT2,dd2ndorder}. Second, in order to enable a certain enhancement mechanism for the magnetic response, the laser fields couple to the medium in such a way that the system becomes  chiral~\cite{chiralindex1,chiralindex2,negindex4}. Unfortunately, an experimental implementation of these schemes to achieve atomic NRI media is very challenging. For example, suitable atomic species are rare, and the required high density implies a number of problems such as Doppler broadening, radiation trapping~\cite{radtrap1,radtrap2}, dephasing or unwanted non-linear processes.

It is important to note, however, that these challenges to a large degree arise from the specific requirements for NRI media. But the two key ingredients, high density and chirality, are of interest in their own rights. This raises the question whether it is possible to study the effects of density or chirality individually at much more accessible parameter ranges. It turns out that the answer is affirmative. Regarding experiments at high density, potential model systems could be, e.g., Rydberg-atoms, doped crystals, or cold gases. In a previous work, we studied dipole-dipole interactions in an electromagnetically induced transparency setup at high density~\cite{myown}. We could show that the polarization contribution to the local field strongly modulates the phase of a weak probe pulse. It was found that this phase modulation distinctively  differs from the nonlinear self-phase modulation experienced by a strong pulse in a Kerr medium.

Here, we turn to the second key ingredient of NRI schemes, and analyze light propagation in a dilute sample of chiral atoms. Quantum optical setups routinely neglect the magnetic field components of the applied electromagnetic fields altogether. Generally, this is justified due to the weak interaction strength, and since magnetic transitions are typically off-resonant from the applied fields. Interestingly, a complete switch from electric to magnetic dipole transitions would not change the obtained results qualitatively, as the structure of the system Hamiltonian is essentially unchanged. It is only the combination of magnetic and electric field interactions in a chiral setup that gives rise to different results. 

We start by solving the wave equations in slowly varying envelope approximation (SVEA) for an arbitrary chiral medium. Our solution allows us to understand the influence of the different response coefficients that characterize a chiral medium on the light propagation. In a next step, we discuss the necessary conditions to implement a chiral atomic medium. Based on these results, we analyze an example system with chiral structure. We study the light propagation through this medium, and find that chiral interactions become relevant already at low medium density. The chirality gives rise to a sensitivity of the system dynamics to the relative phase of the applied fields. We show that this dependence can be used to control the propagation  dynamics of a probe pulse, even throughout the pulse propagation. The results are interpreted based on our analytical solution, and all conclusions are verified using a full numerical solution of the propagation equations.

In a broader context, our results demonstrate that the magnetic field component of the probe field can  crucially influence the propagation dynamics of the light pulse, even though it couples to the medium only very weakly. This may pave the way to further applications of {\it magnetic quantum optics}, utilizing both electric and magnetic components of the applied laser fields. Also, it will turn out that the analysis of chiral media sheds light on the related so-called closed-loop systems~\cite{closed_loop_1,maichen}.  We will find that the chiral media are an ideal implementation of a closed-loop phase-control scheme of light propagation.

The paper is structured as follows. In Sec.~\ref{secchira}, we define and analyze chiral media in terms of parity transformations, and discuss their linear response. The following Sec.~\ref{secgeneral} analyzes light propagation in chiral media. In particular, we analytically solve the chiral wave equations in SVEA. The following Sec.~\ref{secprereq} discusses possible implementations of chiral media in atomic systems. Based on these results, we discuss light propagation in a concrete chiral model system in Sec.~\ref{secmodel}, with the emphasis on phase control of the propagation dynamics. Section~\ref{summary} discusses and summarizes the results.

%%%%%%%%%%%%%%%%%%%%%%%%%%%%%%%%%%%%%%%%%%%%%%%%%%%%%%%%%%%%%%%%%
%%%%%%%%%%%%%%%%%%%%%%%%%%%%%%%%%%%%%%%%%%%%%%%%%%%%%%%%%%%%%%%%%
%%%%%%%%%%%%%%%%%%%%%%%%%%%%%%%%%%%%%%%%%%%%%%%%%%%%%%%%%%%%%%%%%
\section{\label{secchira}Chirality}

An object is defined to be chiral if it is not invariant under parity transformation. Classical electrodynamics is  known to be a parity invariant theory. To see this, and as a basis for our following analysis, we start from the two inhomogeneous macroscopic Maxwell's equations~\cite{jackson},
\begin{subequations}
 \begin{align}
  \nabla \times {\bf E} & = - \partial_t {\bf B}\,,\label{maxeq:1}\\
  \nabla \times {\bf H} & = \partial_t {\bf D}\,,\label{maxeq:2}
 \end{align}
 \label{maxeq}
\end{subequations}
with the electric field ${\bf E}$, the magnetic field ${\bf B}$, the electric displacement field ${\bf D}$, and the magnetizing field ${\bf H}$. The medium is described by its polarization ${\bf P}$ and magnetization ${\bf M}$,
which are related to the quantities entering the Maxwell's equations via the constitutive relations
\begin{subequations}
 \begin{align}
  {\bf D} & = \varepsilon_0 {\bf E} + {\bf P}\,,\label{const:1}\\
  {\bf H} & = \frac{1}{\mu_0} {\bf B} - {\bf M}\,.\label{const:2}
 \end{align}
 \label{const}
\end{subequations}
Due to the used SI unit system, the permittivity $\varepsilon_0$ and the permeability $\mu_0$ of free space occur. We do not consider the two homogeneous Maxwell's equations since with no external sources or currents and assuming transverse fields they are always fulfilled. 

Since ${\bf E}$, ${\bf P}$, and ${\bf D}$ are polar vectors they change sign under a parity transformation. In contrast, ${\bf B}$,  ${\bf M}$ and ${\bf H}$ are unchanged, because they are axial vectors. Together with the sign change of the curl operator, Maxwell's equations Eqs.~(\ref{maxeq}) as well as the constitutive relations Eqs.~(\ref{const}) are parity invariant. 

While the governing equations are parity invariant, this does not need  to be true for the macroscopic medium itself. Assuming a homogeneous isotropic medium, in linear response theory we can define  linear electric and magnetic susceptibilities $\chi_E$ and $\chi_H$,
\begin{subequations}
 \begin{align}
  {\bf P}(t) & = \int \varepsilon_0  \chi_E(\tau) {\bf E}(t - \tau) \textrm{d}\tau \,,\\
  {\bf M}(t) & = \int \chi_H(\tau) {\bf H}(t - \tau) \textrm{d}\tau\,.
 \end{align}
 \label{invlin}
\end{subequations}
These equations only relate quantities of the same, odd or even parity and we see that the coefficients $\chi_E$ and $\chi_H$ are scalars and thus unaffected by a parity transformation. From the perspective of an electrodynamic description, the corresponding medium is not chiral. A different situation arises, if a medium also exhibits cross couplings in a way that an electric field induces magnetization and a magnetizing field induces polarization. In that case Eqs.~(\ref{invlin}) have to be extended by two extra terms,
\begin{subequations}
 \begin{align}
  {\bf P}(t) & = \int \varepsilon_0 \chi_E(\tau) {\bf E}(t - \tau) + \frac{1}{c} \xi_{EH}(\tau) {\bf H}(t - \tau) \textrm{d}\tau ,\\
  {\bf M}(t) & = \int \chi_H(\tau) {\bf H}(t - \tau)  + \frac{1}{c \mu_0}  \xi_{HE}(\tau) {\bf E}(t - \tau) \textrm{d}\tau,
 \end{align}
 \label{chilin}
\end{subequations}
with the cross coupling coefficients $\xi_{EH}$, $\xi_{HE}$, and $c = 1 / \sqrt{\varepsilon_0 \mu_0}$ the speed of light in vacuum. Under a parity transformation, the cross coupling coefficients $\xi_{EH}$ and $\xi_{HE}$ change sign, and thus are pseudo-scalars. If these cross couplings are non-vanishing, the medium properties change under parity transformation and we deal with a chiral medium.

In Eqs.~(\ref{chilin}), we have assumed that the medium and light field configuration are such that the response coefficients are (pseudo-) scalar rather than tensorial as it is the case in general. It turns out that this is possible, e.g., if the probing light field is circularly polarized in the $x-y$ plane and propagates in $\pm z$ direction~\cite{negindex4,negindex5}. This configuration also allows for a different perspective on chirality. A circularly polarized light wave is a chiral object itself, since a left circularly polarized light wave under parity transformation is mapped onto a right circularly polarized light wave and vice versa, rather than onto itself. Because of global parity invariance of electrodynamics, considering a medium interacting with a right circularly polarized light wave is equivalent to considering the inverted medium interacting with a left circularly polarized light wave. It follows that a medium is chiral if its response to a left and right circularly polarized light wave is different. This difference manifests itself in the refractive index, which in a chiral medium with response as in Eq.~(\ref{chilin}) was shown to evaluate to~\cite{negindex4}
\begin{align}
 n_\pm & = \sqrt{ \varepsilon \mu - \frac{1}{4} (\xi_{EH} + \xi_{HE})^2} \mp \frac{i}{2} (\xi_{EH} - \xi_{HE})\,.
 \label{chiind}
\end{align}
Here, $\varepsilon = 1 + \chi_E$ and $\mu = 1 + \chi_H$. The upper sign applies for left circularly polarized light and the lower sign for right circularly polarized light. We see that in Eq.~(\ref{chiind}) it is equivalent to change the sign due to selecting a different polarization or to change the sign of $\xi_{EH}$ and $\xi_{HE}$ due to inverting the medium. Nonzero cross coupling coefficients $\xi_{EH}$ and $\xi_{HE}$ in both cases reflect the chirality of the medium.

%%%%%%%%%%%%%%%%%%%%%%%%%%%%%%%%%%%%%%%%%%%%%%%%%%%%%%%%%%%%%%%%%
%%%%%%%%%%%%%%%%%%%%%%%%%%%%%%%%%%%%%%%%%%%%%%%%%%%%%%%%%%%%%%%%%
%%%%%%%%%%%%%%%%%%%%%%%%%%%%%%%%%%%%%%%%%%%%%%%%%%%%%%%%%%%%%%%%%
\section{\label{secgeneral}Light propagation in chiral media}
In this section, we derive the wave equations for the electric and magnetic field components of a probe field propagating through a chiral medium. This is in contrast to the standard treatment which  assumes a coupling to the electric component only. We apply the slowly varying envelope approximation (SVEA) in order to transform the second order wave equations to equations involving first order derivatives only. Using Fourier transformation techniques, we find the general solution to the first order wave equations, and compare the result to the index of refraction obtained from a single particle susceptibility analysis.

\subsection{\label{secwaveq}Wave equations for the electric and the magnetic field components}

From Eqs.~(\ref{maxeq}) and~(\ref{const}), we can derive wave equations for ${\bf E}$ and ${\bf B}$ with sources described by ${\bf P}$ and ${\bf M}$ as,
\begin{subequations}
 \begin{align}
  \left[ \Delta - \frac{1}{c^2} \partial_t^2 \right] {\bf E} & = \mu_0 \partial_t^2 {\bf P} + \mu_0 \partial_t \nabla \times {\bf M},\\
  \left[ \Delta - \frac{1}{c^2} \partial_t^2 \right] {\bf B} & = \mu_0 \Delta {\bf M} - \mu_0 \partial_t \nabla \times {\bf P}.
 \end{align}
 \label{waveq}
\end{subequations}

In the following, we specialize to the case of a one-dimensional propagation. We choose the positive $z$ axis as the propagation direction and separate a complex envelope function $X_0(z,t)$ from the carrier wave,
\begin{align}
 {\bf X} & = \frac{1}{2} X_0(z,t) {\bf e}_\pm e^{i(k_0 z - \omega_0 t)} + \textrm{c.c.},
\end{align}
where ${\bf X}$ stands for ${\bf E}$, ${\bf B}$, ${\bf P}$, or ${\bf M}$. The complex unit vector for left or right circular polarized light is ${\bf e}_\pm$, and the wave number of the carrier wave in vacuum is $k_0$ with $\omega_0 = c k_0$ as the corresponding frequency. Assuming that variations in space and time of the envelope function are on a much larger scale than the wavelength $\lambda_0 = 2 \pi / k_0$ and the oscillation period $T_0 = 2 \pi / \omega_0$ of the carrier wave, we can apply the slowly varying envelope approximation (SVEA)~\cite{scullybook}. In this approximation, derivatives of the envelope function in space and time are neglected compared to derivatives of the carrier wave. As a consequence, SVEA reduces the wave equations Eqs.~(\ref{waveq}) to first order equations for the envelope functions,
\begin{subequations}
 \begin{align}
  \left[ \partial_z + \frac{1}{c} \partial_t \right] E_0(z,t) & = \frac{i k_0}{2 \varepsilon_0} P_0(z,t) \mp \frac{k_0}{2 \varepsilon_0 c} M_0(z,t)\,,\\
  \left[ \partial_z + \frac{1}{c} \partial_t \right] B_0(z,t) & = \frac{i k_0 \mu_0}{2} M_0(z,t) \pm \frac{k_0}{2 \varepsilon_0 c} P_0(z,t)\,.
 \end{align}
 \label{sveaq}
\end{subequations}
As before and in the following, the upper sign applies for left circular polarization whereas the lower sign applies for right circular polarization. Eqs.~(\ref{sveaq}) are considerably simpler to solve analytically as well as numerically than the second order Eqs.~(\ref{waveq}) we derived initially. Yet, they still incorporate the essential physics of light propagation in one spatial dimension for a chiral medium with a coupling to both the electric and the magnetic component of a weak probe field.

\subsection{\label{secsveas}Solution of the wave equations}
We will now solve the wave equations (\ref{sveaq}) in SVEA to a level that enables us to determine how the actual propagation dynamics depends on the four quantities $\chi_E$, $\chi_H$, $\xi_{EH}$, and $\xi_{HE}$ characterizing the medium. First, we transform Eqs.~(\ref{sveaq}) into Fourier space. To substitute $P_0$ and $M_0$, we also Fourier transform Eqs.~(\ref{chilin}). Next, to replace $H_0$ by $B_0$, we use the Fourier transform of Eq.~(\ref{const:2}). Finally, we can combine both wave equations into a single matrix equation for the vector ${\bf F}(z,\Delta_p)=(E_0(z,\Delta_p),c B_0(z,\Delta_p))^T$,
\begin{align}
 \partial_z\,  {\bf F}(z,\Delta_p) 
  = i k_0\,  \mathcal{M} \, {\bf F}(0,\Delta_p) \,,
 \label{matriq}
\end{align}
and the elements of the matrix $\mathcal M$ are given by
\begin{subequations}
\begin{align}
\mathcal{M}_{1,1} &= \frac{\Delta_p}{\omega_o} + \frac{1}{2 \mu} (\chi_E - \xi_{EH} \xi_{HE} \pm i \xi_{HE}) \,,\\
\mathcal{M}_{1,2} &= \frac{1}{2 \mu} (\xi_{EH} \pm i \chi_H)\,,\\
\mathcal{M}_{2,1} &=\frac{1}{2 \mu}(\xi_{HE} \mp i \mu \chi_E \pm i \xi_{EH} \xi_{HE}) \,,\\
\mathcal{M}_{2,2} &= \frac{\Delta_p}{\omega_0} + \frac{1}{2 \mu} (\chi_H \pm i \xi_{EH})\,.
\end{align}
\end{subequations}
In these equations, the detuning $\Delta_p = \omega - \omega_0$ accounts for the frequency distribution of the envelope functions around the carrier frequency. Applying the vacuum phase relation for circularly polarized light, $c B_0 = \mp i E_0$, we find that the initial condition for the evolution Eq.~(\ref{matriq}),
\begin{align}
 {\bf F}(0, \Delta_p) = 
 \begin{pmatrix}
  E_0(0, \Delta_p)\\
  \mp i E_0(0, \Delta_p)\\
 \end{pmatrix}
 \,,
\end{align}
 is an eigenvector to matrix $\mathcal{M}$ in Eq.~(\ref{matriq}). Therefore, the solution can be calculated with the help of the corresponding eigenvalue $\eta$. The solution reads
\begin{align}
 {\bf F}(z, \Delta_p)
  = e^{i k_0 \eta z}
 \begin{pmatrix}
  E_0(0, \Delta_p)\\
  \mp i E_0(0, \Delta_p)\\
 \end{pmatrix}
 \,,
 \label{solwav}
\end{align}
and the frequency dependence is contained in the eigenvalue
\begin{align}
 \eta = \frac{\Delta_p}{\omega_0} + \frac{1}{2\mu} \left[\chi_E\, \mu  + \chi_H - \xi_{EH} \xi_{HE} \mp i (\xi_{EH} - \xi_{HE})\right]\,.
 \label{eta}
\end{align}
This result can be compared to the refractive index Eq.~(\ref{chiind}) in a chiral medium obtained from a single-particle susceptibility analysis. The application of SVEA is equivalent to the condition $|n_\pm - 1| \ll 1$, such that we can perform a first order Taylor expansion of the square root contribution in Eq.~(\ref{chiind}). Keeping terms linear in the response coefficients we find
\begin{align}
 n_\pm = 1 + \frac{1}{2} (\chi_E + \chi_H) \mp \frac{i}{2} (\xi_{EH} - \xi_{HE})\,.
 \label{sveaind}
\end{align}
Linearizing Eq.~(\ref{eta}) and taking into account the contribution of the carrier wave, we find the same result as in Eq.~(\ref{sveaind}). This demonstrates the consistency of our solution with the previously calculated chiral index of refraction. Our result based on the solution of the wave equations, however, has the advantage that it offers the possibility to study propagation dynamics. Because of the simple formulas we have derived, analytical solutions become accessible. If for a specific system the frequency dependencies of $\chi_E$, $\chi_H$, $\xi_{EH}$, and $\xi_{HE}$ are known, they can be used together with the solution Eq.~(\ref{solwav}) in frequency space  to calculate the actual dynamics in the time domain.

%%%%%%%%%%%%%%%%%%%%%%%%%%%%%%%%%%%%%%%%%%%%%%%%%%%%%%%%%%%%%%%%%
%%%%%%%%%%%%%%%%%%%%%%%%%%%%%%%%%%%%%%%%%%%%%%%%%%%%%%%%%%%%%%%%%
%%%%%%%%%%%%%%%%%%%%%%%%%%%%%%%%%%%%%%%%%%%%%%%%%%%%%%%%%%%%%%%%%
\section{\label{secprereq}Chiral atomic media}

In this section we discuss the general conditions for realizing a chiral atomic medium in the form as introduced in Sec.~\ref{secchira}. As a first condition, a medium in addition to an electric probe field transition should exhibit a near-degenerate transition coupling to the magnetic component of the probe field. A second condition is that, to enable chiral cross couplings, there must be a mechanism to allow for polarization [magnetization] to be induced by the magnetic [electric] probe field component.

The first condition can be difficult to fulfill in atomic media. A small frequency difference between electric and magnetic probe field transition could be removed by an external magnetic field leading to a Zeeman shift, but magnetic dipole transitions are usually found in a much smaller frequency range compared to electric dipole transitions. The reason is that due to selection rules imposed by the corresponding transition matrix element, magnetic dipole transitions can only occur between states of the same parity. This implies that magnetic dipole transitions only take place between states with the same angular momentum quantum number whereas electric dipole transitions connect states of different angular momentum. A possible solution are atomic species with a large spin-orbit coupling~\cite{haken}. Then, states with the same angular momentum quantum number can display a large splitting and their transition frequency can become comparable to the transition frequency of an electric dipole transition. This could also help to realize a probe field transition frequency in the optical domain which is especially attractive for applications. Motivated by ideas like NRI, several atomic species have been investigated with respect to this condition. Dysprosium~\cite{negindex4,dyspec}, Hydrogen~\cite{negindex2}, metastable Neon and other noble gases~\cite{negindex2,porthdip} have been identified as promising candidates. An alternative approach could be to make use of non-resonant two-photon transitions in order to relax the requirement of having both electric and magnetic transitions at similar transition frequencies~\cite{negindex5,porthdip}.

The second condition can be implemented in closed-loop systems. In these systems, the laser-driven transitions form a closed interaction contour such that photon emission and absorption can take place in a cycle~\cite{closed_loop_1,closed_loop_2}. Because the closed interaction contour enables quantum mechanical pathway interference, it can render the system phase dependent. Closed-loop systems have mainly been investigated in non-chiral contexts~\cite{maichen,korsunsky,morigi,malinovsky,windholz,shpaism,kajari}. Here, the closed interaction contour only includes the coupling to one (electric) probe field component. This is known to induce extra contributions to the medium response, but these cannot be utilized for light propagation of a pulsed probe field in a straightforward way~\cite{closed_loop_3,closed_loop_4}. The reason is that already without the probe field, such a configuration would give rise to parametric processes which  scatter the control fields into the probe field mode~\cite{closed_loop_3,closed_loop_4,veer,merriam,hinze}. As we will see in Sec.~\ref{secmodel}, this conclusion changes in the case of chiral closed-loop systems. There, we will find that the closed-loop phase can provide a convenient control parameter for light propagation in chiral media.

%%%%%%%%%%%%%%%%%%%%%%%%%%%%%%%%%%%%%%%%%%%%%%%%%%%%%%%%%%%%%%%%%
%%%%%%%%%%%%%%%%%%%%%%%%%%%%%%%%%%%%%%%%%%%%%%%%%%%%%%%%%%%%%%%%%
%%%%%%%%%%%%%%%%%%%%%%%%%%%%%%%%%%%%%%%%%%%%%%%%%%%%%%%%%%%%%%%%%
\section{\label{secmodel}Chiral control of propagation dynamics}

\subsection{\label{subsecmodel}Model system}
\begin{figure}[t]
\includegraphics[width=6cm]{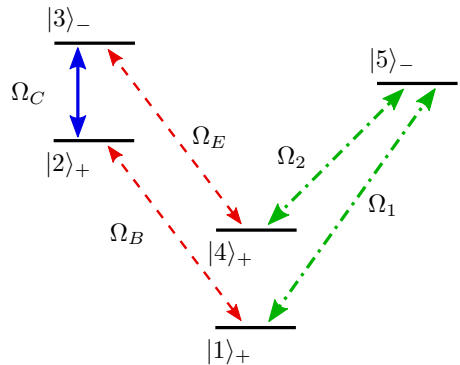}
\caption{\label{fig1}(Color online) Atomic level scheme used to describe the chiral medium. Two strong control fields with Rabi frequencies $\Omega_1$ and $\Omega_2$ prepare a coherent superposition of states $\ket{1}$ and $\ket{4}$. The magnetic and electric probe beam components couple to two transitions with Rabi frequencies $\Omega_B$ and $\Omega_E$, respectively. $\Omega_C$ describes an additional control field. The relative state parities are denoted by $+$ or $-$.}
\end{figure}

In this Section, we apply the results obtained so far to discuss light propagation in a specific system with simultaneous coupling to the electric and the magnetic component of a probe laser field. The atomic medium is modeled as the five level system shown in Fig.~\ref{fig1}. It was first introduced in the context of generating atomic media exhibiting a negative index of refraction~\cite{negindex4}. 
In zeroth order of the probe field, the atomic medium forms an effective three-level system in $\Lambda$-configuration consisting of states $|1\rangle, |4\rangle$ and $|5\rangle$. Two strong resonant control fields with Rabi frequencies $\Omega_1$ and $\Omega_2$ prepare the atoms in a dark state, i.e., a stable coherent superposition of states $\ket{1}$ and $\ket{4}$.
The magnetic and electric probe field components couple to degenerate magnetic and electric dipole transitions with Rabi frequencies $\Omega_B$ and $\Omega_E$, and the upper levels of these two dipole transitions are coupled by an additional resonant control field with Rabi frequency $\Omega_C$. We assume that state $\ket{1}$ and $\ket{4}$ have the same parity. Because the electric probe field transition couples states of different parity whereas the magnetic probe field transition couples states of the same parity, the additional control field couples to an electric dipole transition. Note that the Rabi frequencies $\Omega_i=|\Omega_i|\,\exp[i(\vec{k}_i\vec{r}+\varphi_i)]$ are complex, and contain the wave vector $\vec{k}_i$ and the absolute phase $\varphi_i$ ($i\in\{1,2,C,E,B\}$).

As discussed in Sec.~\ref{secintro}, the linear response of the chiral atomic medium is characterized by two contributions.  First, direct contributions arise from the polarization [magnetization] induced by the electric [magnetic] probe field component.  It turns out that the direct contribution to the magnetic response is typically small. The reason is that magnetic dipole couplings usually are much weaker than electric dipole couplings,
\begin{align}
 m/d \sim \alpha c \quad \Rightarrow \quad |\Omega_B| \sim \alpha |\Omega_E|\,,
\end{align}
with magnetic dipole moment $m$, electric dipole moment $d$, and fine structure constant $\alpha$. Therefore, in comparison to the direct electric response, the direct magnetic response is parametrically suppressed by two powers of the fine structure constant~\cite{bastian},
\begin{align}
 \chi_H \sim \alpha^2 \chi_E\,.
\end{align}
The other contributions to the medium response are cross couplings, which arise due to polarization [magnetization] induced by the magnetic [electric] probe field component. It was found that the considered level scheme displays a relevant magnetic response due to these cross contributions~\cite{negindex4,bastian}.  A magnetic response arising from such a cross coupling is only  suppressed by a factor on the order of $\alpha$. Furthermore, it is proportional to the strength of the ground state coherence of state $\ket{1}$ and $\ket{4}$ and the control field $\Omega_C$. Since cross couplings of this type lead to nonzero values for $\xi_{EH}$ and $\xi_{HE}$, they also render the medium chiral.

The Hamiltonian of the system in Fig.~\ref{fig1} in a suitable interaction picture is given by~\cite{scullybook}
\begin{align}
\label{ham}
H_I = & -\hbar \Delta_p (A_{22}+A_{33}) \nonumber \\
& -\frac{\hbar}{2}\Bigl (\Omega_B\, A_{21} + \Omega_E\, A_{34} + \Omega_C\, A_{32} \nonumber \\
& + \Omega_1\, A_{51} + \Omega_2\, A_{54}+ H.c. \Bigr)
\,,
\end{align}
where $\Delta_p$ is the probe field detuning and atomic transition operator is defined as $A_{jk} = \ket{j}\bra{k}$. An interesting insight into the involved physics can be gained by applying a further unitary transformation to the Hamiltonian
\begin{align}
\label{ham-cl}
\bar{H_I} = & -\hbar \Delta_p (A_{22}+A_{33}) \nonumber \\
& -\frac{\hbar}{2}\Bigl (|\Omega_B|\, A_{21} + |\Omega_E|\, A_{34} + |\Omega_C|\,e^{i\Phi} A_{32} \nonumber \\
& + |\Omega_1|\, A_{51} + |\Omega_2|\, A_{54} + H.c. \Bigr)
\,.
\end{align}
It can be seen that $H_I$ contains a phase contribution
\begin{align}
\label{phi}
\Phi = (\varphi_2-\varphi_1+\varphi_c) + (\vec{k}_2-\vec{k}_1+\vec{k}_C)\vec{r}\,,
\end{align}
which cannot be removed entirely via unitary transformations.  The origin of this phase dependence is the closed-loop structure of the considered level scheme. Starting from $\ket{1}$, the system can evolve in a non-trivial loop sequence $\ket{1} \to \ket{5} \to \ket{4} \to \ket{3} \to \ket{2} \to \ket{1}$ back to the initial state. This enables pathway interference, e.g., from $\ket{1}$ to $\ket{3}$ either via $\ket{5}$ and $\ket{4}$, or via $\ket{2}$. The phase difference between these two interfering pathways is equivalent to the phase difference given by Eqs.~(\ref{ham-cl}) and (\ref{phi}). It is interesting to note that the closed-loop phase $\Phi$  does not contain properties of the probe field, which is due to the fact that a closed transition loop involves both an absorption and an emission of a probe field photon. Furthermore, $\Phi$ does not depend on time, because of the cancellation of the probe field and because the three control fields $\Omega_1, \Omega_2, \Omega_C$ are applied on resonance. 

We are now also in the position to understand why a phase-control of light propagation is possible in chiral closed-loop media, in contrast to closed-loop media coupling only to the electric component of the probe field. From Fig.~\ref{fig1}, it can be seen that in the presence of the probe field, a closed loop is established, such that the medium becomes sensitive to the closed-loop phase $\Phi$. But in the absence of the probe field, two transitions in the closed loop structure are undriven.  It follows that parametric processes scattering the control fields into the probe field mode cannot occur, independent of the phase matching condition assumed for the wave vectors of the probe and driving fields. This crucial difference to closed-loop media coupling to one (electric) probe field component only enables the phase control of light propagation that we will find in Sec.~\ref{secphase}.

%%%%%%%%%%%%%%%%%%%%%%%%
%%%%%%%%%%%%%%%%%%%%%%%%
%%%%%%%%%%%%%%%%%%%%%%%%
%%%%%%%%%%%%%%%%%%%%%%%%

In the following, we continue our analysis based on the Hamiltonian $H_I$ in Eq.~(\ref{ham}) and assume that the laser configuration satisfies the usual phase matching condition $\vec{k}_2-\vec{k}_1+\vec{k}_C$, such that the system dynamics only depends on the relative phase 
\begin{align}
\label{phase}
\Phi_0 = \varphi_2-\varphi_1+\varphi_c\,.
\end{align}
In zeroth order of the probe fields, the equations of motion (EOM) governing the dynamics of the atomic degrees of freedom can easily be solved. The non-vanishing density matrix elements are
\begin{subequations}
\begin{align}
 \rho_{11}^{(0)} & = \frac{|\Omega_2|^2}{|\Omega_1|^2 + |\Omega_2|^2}\,,\\
 \rho_{44}^{(0)} & = \frac{|\Omega_1|^2}{|\Omega_1|^2 + |\Omega_2|^2}\,,\\
 \rho_{41}^{(0)} & = - \frac{\Omega_1 \Omega_2^*}{|\Omega_1|^2 + |\Omega_2|^2}\,.
\end{align}
\end{subequations}
The linear response to a weak probe field does not disturb this dark state prepared by $\Omega_1$ and $\Omega_2$, and can thus be calculated on the basis of the lowest order populations.
In order to make connection to NRI calculations, we also take into account local field corrections (LFC) as they occur in denser gases, even though we will find later that magnetic field effects can already be observed at densities low enough to neglect LFC. In the literature, a generalized form of the Clausius-Mossotti equation~\cite{jackson} has been used to numerically extract $\chi_E$, $\chi_H$, $\xi_{EH}$, and $\xi_{HE}$ from the steady state solution to linear order in the local fields $E_\textrm{loc}$ and $B_\textrm{loc}$ in~\cite{negindex4}. We pursue a different strategy here which provides us with an analytic solution including LFC~\cite{myown}.  We directly replace the local fields in the EOM by the external fields $E_\textrm{ext}$ and $B_\textrm{ext}$ with the help of the Lorenz-Lorentz relations~\cite{jackson}
\begin{subequations}
\begin{align}
 E_\textrm{loc} & = E_\textrm{ext} + \frac{1}{3 \varepsilon_0} P\,,\\
 B_\textrm{loc} & = B_\textrm{ext} + \frac{\mu_0}{3} M\,.
\end{align}
\end{subequations}
This leads to nonlinear EOM if the polarization $P$ and the magnetization $M$ are expressed by microscopic expectation values of the electric and magnetic dipole density containing elements of the atomic density matrix. Since we are interested in linear response, we expand the  EOM up to linear order in the probe field, and solve for the steady state. From this steady state solution, we find for the response coefficients
\begin{subequations}
\begin{align}
 \chi_E = & \frac{3 L \gamma_{34} \rho_{44}^{(0)} \frac{i}{2} \Gamma_{24}}{\Gamma_{34} \Gamma_{24} + \frac{|\Omega_C|^2}{4}},\\
 \chi_H = & \frac{3 L \gamma_{34} \alpha^2 \rho_{11}^{(0)} \frac{i}{2} \Gamma_{31}}{\Gamma_{21} \Gamma_{31} + \frac{|\Omega_C|^2}{4}}
   \Bigl(1 - \frac{\xi_{EH} |\rho_{41}^{(0)}|\, \frac{|\Omega_C|}{4}\,e^{-i\Phi_0}}{3 \alpha \, \rho_{11}^{(0)}\, \frac{i}{2} \Gamma_{31}} \Bigr) ,\\
  \xi_{EH} = & \frac{3 L \gamma_{34} \alpha \, |\rho_{41}^{(0)}|\, \frac{|\Omega_C|}{4}}{\Gamma_{34} \Gamma_{24} + \frac{|\Omega_C|^2}{4}}\,e^{i\Phi_0},\\
  \xi_{HE} = & \frac{3 L \gamma_{34} \alpha\, |\rho_{41}^{(0)}|\, \frac{|\Omega_C|}{4}}{\Gamma_{21} \Gamma_{31} + \frac{|\Omega_C|^2}{4}}
  \left(1 + \frac{\chi_E}{3} \right)\,e^{-i\Phi_0},
\end{align}
\end{subequations}
where $L = N \lambda^3/ 4 \pi^2$ includes the dependence on the number of particles $N\lambda^3$ per wavelength $\lambda$ cubed. The decay rate and the detuning including the typical LFC shift for the electric and magnetic probe field transitions are given by
\begin{subequations}
\begin{align}
\Gamma_{34} &= \gamma / 2 + \gamma_\textrm{dec} - i(\Delta_p + \rho_{44}^{(0)} L \gamma_{34} / 2)\,,\\
\Gamma_{21} &= \gamma_{21} / 2 + \gamma_\textrm{dec} - i \Delta_p\,,
\end{align}
\end{subequations}
where $\gamma = \gamma_{31} + \gamma_{32} + \gamma_{34}$ is the overall decay rate of state $\ket{3}$, the linewidth of the magnetic probe field transition is $\gamma_{21}$,  and $\gamma_{dec}$ accounts for additional decoherence induced by atomic collisions. 

Interpreting the dark state of $\ket{1}$ and $\ket{4}$ as a single quantum state characterized by the coherence between the two states, both probe field transitions can be viewed as being part of a three level electromagnetically induced transparency (EIT) system~\cite{eit1, eit2}. The electric probe field and the control field $\Omega_C$ form a $\Lambda$-system for which 
\begin{align}
\Gamma_{24} = \gamma_{21} / 2 - i \Delta_p 
\end{align}
contains the ground state decoherence and the two-photon detuning. No additional decoherence $\gamma_\textrm{dec}$ besides the decay of the magnetic probe field transition has to be included if we assume that levels $\ket{2}$ and $\ket{4}$ are nearly degenerate, as then phase-perturbing collisions in the medium lead to random, but correlated phase shifts of the two states~\cite{negindex4}. 
The magnetic probe field and the control field $\Omega_C$ form a ladder system in which the decoherence and two-photon detuning relevant for EIT is described by 
\begin{align}
\Gamma_{31} = \gamma / 2 + 2 \gamma_\textrm{dec} - i \Delta_p \,.
\end{align}

By inspection of the structure of the solution for $\chi_E$ and $\chi_H$, we find that the direct electric and magnetic responses of the EIT system are amended by LFC. The electric response features the usual EIT transparency window which is only degraded by the weak ground state decoherence induced by $\gamma_{21}$. The transparency window of the direct magnetic response on the other hand is not as prominent due to strong decoherence $\gamma / 2 + 2 \gamma_\textrm{dec}$ in the ladder system. But this does not affect the propagation dynamics much since $\chi_H$ is suppressed by a factor $\alpha^2$ compared to $\chi_E$ as discussed above. In particular for propagation dynamics at densities corresponding to $L \le 1$, it can be neglected. On the contrary, the chiral cross couplings are only suppressed by a factor of $\alpha$ and do influence the propagation dynamics. As expected, they are proportional to $|\rho_{41}^{(0)}|$, $|\Omega_C|$, and depend on the relative closed-loop phase $\Phi_0$. At first glance it might look like a phase dependence is also induced to $\chi_H$ by LFC because it depends on $\Phi_0$. But a closer look reveals that this phase dependence is exactly canceled by the factor $\xi_{EH}$. From these solutions we expect a propagation with a reduced group velocity and with a phase dependent refractive index induced by the chiral cross couplings.
%%%%%%%%%%%%%%%%%%%%%%%%
%%%%%%%%%%%%%%%%%%%%%%%%
\begin{figure}[t]
\center
\includegraphics[width=0.9\columnwidth]{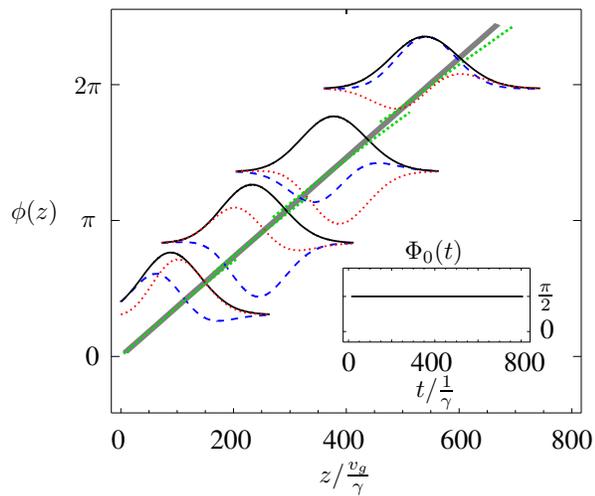}
\caption{\label{fig2}(color online) 
Propagation dynamics in a chiral medium. The closed-loop phase $\Phi_0 = \pi/2$ is kept constant, see inset. The absolute value (solid black lines), the real part (dashed blue lines), and the imaginary part (dotted red lines) of the probe pulse envelope are plotted for successive propagation times. Arbitrary, but consistent units are used and successive plots are vertically offset by the overall phase at the center of the pulse. The spatial phase dependence of each pulse (dashed green straight lines) is compared to the value expected from the analytical solution (thick gray straight line).}
\end{figure}
%%%%%%%%%%%%%%%%%%%%%%%%
%%%%%%%%%%%%%%%%%%%%%%%%

One further observation about the chirality of the medium can be made. In the sense of a macroscopic electrodynamic description, the control fields $\Omega_C$, $\Omega_1$, and $\Omega_2$ are part of the medium. Thus, they change sign when the medium is inverted and lead to the expected chiral behavior of $\xi_{EH}$ and $\xi_{HE}$ as described in Sec.~\ref{secchira}. 

With the analytic solution for $\chi_E$, $\chi_H$, $\xi_{EH}$, and $\xi_{HE}$ at hand, we can now proceed to study propagation effects in a chiral medium, focusing on control of the propagation dynamics via the closed-loop phase as concrete application.

\subsection{\label{secphase}Phase switching}
In this Section we study the propagation dynamics of a pulsed probe field in the chiral medium discussed in Sec.~\ref{subsecmodel}. Due to the direct electric response, the probe pulse propagates with a reduced group velocity, as it is the case in a usual EIT system. If we write the group velocity
\begin{align}
 v_g = \frac{c}{1 + n_g},
 \label{vg}
\end{align}
the group index $n_g$ it is given by
\begin{align}
 n_g = \frac{3 L \omega \gamma_{34} \rho_{44}^{(0)}}{|\Omega_C|^2}.
 \label{ng}
\end{align}
This is the standard expression in a EIT system, except for the scaling by the population fraction $\rho_{44}^{(0)}$ in the ground state of the electric probe field transition. The direct magnetic response can be neglected and all additional influence of the medium on a probe pulse is due to the chiral cross coupling of the electric and magnetic probe field components. 

The chiral cross coupling is mediated by the closed interaction loop and depends on the relative phase $\Phi_0$ between the three control fields. Since the probe field propagates with strongly reduced group velocity whereas the control field $\Omega_C$ propagates at speed of light, a phase change of $\Phi_0$ during the propagation of the probe pulse changes the  frequency dependent refractive index for the probe field nearly instantaneously. In a similar fashion as a light pulse can be slowed down, stored, and retrieved by changing the amplitude of the control field~\cite{stopedlight}, the phase of the control field can be used as a tool to control the refractive index for the probe pulse.

To clarify this idea further, in the following, we present two exemplary results from a numerical solution of the atomic EOM coupled to the wave equations for the two probe field components and the control field. We assume a medium with a density corresponding to $L = 0.01$. For $\lambda = 795$nm, this translates to $N \sim 8 \times 10^{11} \textrm{cm}^{-3}$, which is low enough to neglect LFC. The control field strength is chosen as  $|\Omega_C |= 2 \gamma$, and the dark state as symmetric, $|\Omega_1| = |\Omega_2|$, such that $\rho_{11}^{(0)} = \rho_{44}^{(0)} = 0.5$ and $\rho_{41}^{(0)} = -0.5$. In both cases, we propagate a Gaussian probe pulse with a width in time of $\sigma = 50/\gamma$. Its spectrum is well contained in the EIT transparency window such that it suffers very little absorption and broadening. We use a collisional decoherence of $\gamma_{dec} = 0.5 \gamma$ and for the probe field we assume left circular polarization.

%%%%%%%%%%%%%%%%%%%%%%%%
%%%%%%%%%%%%%%%%%%%%%%%%
\begin{figure}[t]
\center
\includegraphics[width=0.9\columnwidth]{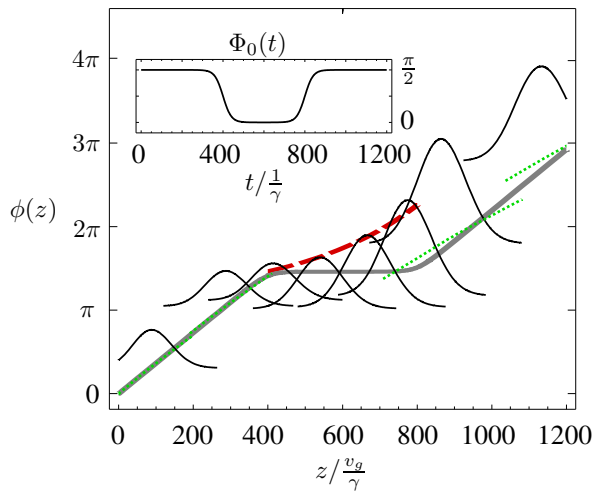}
\caption{\label{fig3}(color online) Propagation dynamics in a chiral medium with phase control. The closed-loop phase is switched from $\pi/2$ to zero and back to $\pi/2$ throughout the propagation, as shown in the inset. The absolute value (black solid lines) of the probe pulse envelope is plotted for successive propagation times. The scaling and offsets are used as in Fig.~\ref{fig2}. At the beginning and at the end of the propagation the spatial phase dependence of each pulse (dashed green straight lines) is compared to the value expected from the analytical solution (thick gray line). During the time period of zero control phase, probe pulse envelope phase does not change and the pulse experiences gain instead, as indicated by the thick dashed red line.}
\end{figure}
%%%%%%%%%%%%%%%%%%%%%%%%
%%%%%%%%%%%%%%%%%%%%%%%%

In the first example, the phase $\Phi_0$ is chosen constant as $\pi/2$.  In Fig.~\ref{fig2}, we show the corresponding propagation dynamics. The envelope of the electric probe field component is depicted for four successive propagation times, together with the real and imaginary part of the envelope and the pulse phase. The group velocity observed in the numerical calculation is consistent with Eqs.~(\ref{vg}) and~(\ref{ng}). We find that throughout the propagation, the phase of the pulse  increases linearly. Thus, the real and imaginary parts of the envelope interchange while the absolute value of the envelope remains constant. This dynamics is due to a positive real index of refraction at resonance. Its theoretical value can be deduced from the solutions given in Secs.~\ref{secsveas} and~\ref{secmodel}. Neglecting $\chi_H$, the eigenvalue Eq.~(\ref{eta}) reduces to
\begin{align}
 \eta = \frac{\Delta_p}{\omega_0} + \frac{1}{2} \left [\chi_E \mp i (\xi_{EH} - \xi_{HE})\right ]
\,.
\end{align}
The term linear in $\Delta_p$ arising from $\chi_E$ leads to the reduced group velocity given  in Eq.~(\ref{vg}). We approximate $\xi_{EH}$ and $\xi_{HE}$ by the constant term at resonance leading to a constant refractive index. Fourier transforming Eq.~(\ref{solwav}) back into time domain, we find
\begin{align}
 \begin{pmatrix}
  \Omega_E(z, t)\\
  \Omega_B(z, t)\\
 \end{pmatrix}
  =
 \begin{pmatrix}
  \Omega_E \left(0, t - z/v_g\right)\\
  \Omega_B \left(0, t - z/v_g\right)\\
 \end{pmatrix}
  \,e^{i \beta k_0 z}\,,
\end{align}
in which the polarization-dependent $\beta$  is given by 
\begin{align}
 \beta = \pm i \alpha \frac{n_g \rho_{41}^{(0)} |\Omega_C|}{2 \omega \rho_{44}^{(0)}} 
\left(e^{i\Phi_0} - \frac{|\Omega_C|^2 e^{-i\Phi_0}}{2 \gamma_\textrm{dec} \gamma + 8 \gamma_\textrm{dec}^2 + |\Omega_C|^2} \right)\,,
 \label{beta}
\end{align}
where we neglected the radiative decay of the magnetic probe field transition $\gamma_{21}$ which is small compared to $\gamma_{dec}$.

Inserting the parameters chosen in Fig.~\ref{fig2}, we obtain a positive real value of $\beta$ such that the phase of the envelope increases constantly with propagation distance. In Fig.~\ref{fig2}, we compare this phase dependence obtained from our approximate analytical model with the phase from the numerical propagation of the pulse at different propagation distances. At the center of the pulse the agreement between numerical and analytical solution is very good. Away from the center there are small deviations that increase with propagation distance. These deviations can be attributed to the neglected frequency dependence of $\xi_{EH}$ and $\xi_{HE}$.

%%%%%%%%%%%%%%%%%%%%%%%%
%%%%%%%%%%%%%%%%%%%%%%%%
\begin{figure}[t]
\center
\includegraphics[width=0.9\columnwidth]{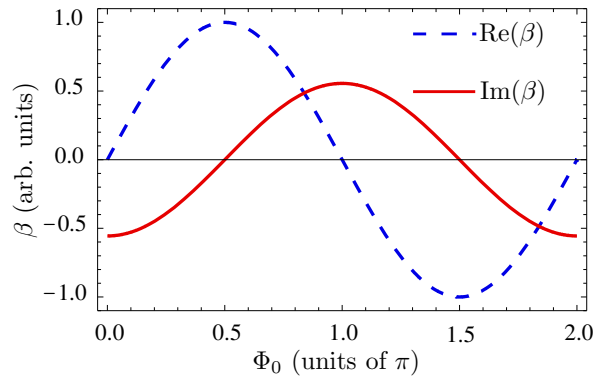}
\caption{\label{fig4}(color online) Real (dashed blue line) and imaginary (solid red line) part of $\beta$ against the closed-loop phase $\Phi_0$.}
\end{figure}
%%%%%%%%%%%%%%%%%%%%%%%%
%%%%%%%%%%%%%%%%%%%%%%%%

In a second example, the phase $\Phi_0$ is changed during the propagation. Initially, it is set to $\pi/2$ as in the first example. During the course of propagation, it is switched to zero for an intermediate period of time and finally, it is switched back to $\pi/2$. The numerical results for the corresponding propagation dynamics are shown in Fig.~\ref{fig3}, and the time-dependent value of the phase is shown as inset in Fig.~\ref{fig3}.  In the initial and final part of the propagation with phase $\pi/2$, we find the same propagation dynamics as in Fig.~\ref{fig2}, with reduced group velocity $v_g$ and with a linear increase of the phase. But during the period with control phase switched to zero, the phase of the pulse envelope remains approximately constant. Instead, its amplitude increases exponentially. As in the first example, this dynamics can be explained based on the index of refraction at resonance. Calculating $\beta$ for $\Phi_0 = 0$ in the analytical solution, we find a negative imaginary value corresponding to gain. Again, we compare the analytical solution to the numerical data. The phase dependence is given by a linear increase with a plateau in between, whereas the increase in amplitude follows an exponential function, see Fig.~\ref{fig3}. Both analytical solutions agree reasonably well with the numerical data, given that the finite switch period with the corresponding transient dynamics is not taken into account in the analytical calculation.

These examples show that by changing the relative control field phase $\Phi_0$, the dynamics of the probe field can be substantially influenced throughout the propagation. In the second example, the refractive index experienced by a pulsed probe field with left circular polarization was changed from a positive real value $n \ge 1$ to an imaginary value representing gain and back. With different values for the control phase, also $n \le 1$, absorption, or intermediate cases are possible. This is illustrated in Fig.~\ref{fig4}, which depicts the real and the imaginary part of $\beta$ obtained from our analytical calculation against the control phase. 
%

%%%%%%%%%%%%%%%%%%%%%%%%%%%%%%%%%%%%%%%%%%%%%%%%%%%%%%%%%%%%%%%%%
%%%%%%%%%%%%%%%%%%%%%%%%%%%%%%%%%%%%%%%%%%%%%%%%%%%%%%%%%%%%%%%%%
%%%%%%%%%%%%%%%%%%%%%%%%%%%%%%%%%%%%%%%%%%%%%%%%%%%%%%%%%%%%%%%%%
\section{\label{summary}Discussion and summary}
In summary, we discussed light propagation dynamics in a chiral atomic medium. The chirality arises from a cross coupling of the electric and the magnetic component of a probe field induced by the medium. First, we derived wave equations for both probe field components in the chiral medium. Using the slowly varying envelope approximation, we then solved these equations in Fourier space. Our solution elucidates the actual dependence of the propagation dynamics on the direct and chiral response functions in a macroscopic electrodynamic description. Based on these results, we discussed the general conditions necessary for realizing a chiral atomic medium. In a concrete example, we analytically determined the response functions for a specific chiral model system from the microscopic quantum mechanical equations of motion. As our main result, we demonstrated that chiral effects can crucially influence the propagation dynamics already at experimentally accessible parameter ranges. In particular, we find that chiral effects can already be observed in dilute vapors. As an application, we showed that the chiral couplings render the medium sensitive to the relative phase of the applied fields. By changing the relative phase of one of the applied coupling fields, the propagation of a probe pulse can be controlled dynamically during its passage through the medium. From this, we concluded that chiral media are an ideal implementation of closed-loop phase control of light propagation. In a broader context, it is remarkable that the magnetic field component of the probe field can substantially modify the probe pulse propagation dynamics, despite the weak coupling of the magnetic field to the medium. It remains to be seen if such magnetic field couplings may also find other applications in quantum optics.

%%%%%%%%%%%%%%%%%%%%%%%%%%%%%%%%%%%%%%%%%%%%%%%%%%%%%%%%%%%%%%%%%
%%%%%%%%%%%%%%%%%%%%%%%%%%%%%%%%%%%%%%%%%%%%%%%%%%%%%%%%%%%%%%%%%
%%%%%%%%%%%%%%%%%%%%%%%%%%%%%%%%%%%%%%%%%%%%%%%%%%%%%%%%%%%%%%%%%

\end{document}